\documentclass[aps,prb,floatfix,showpacs,twocolumn]{revtex4}
\usepackage{graphicx}
\begin{document}

\title{Phase separation of a model binary polymer solution in an external
field.}

\author{Chris I. Addison$\dagger$, Pierre-Arnaud Artola$\dagger$$\ddagger$,
Jean-Pierre Hansen$\dagger$ and Ard A. Louis$\dagger$}

\address{$\dagger$Department of Chemistry, University of Cambridge, CB2 1EW
(UK)\\ $\ddagger$Permanent address: Ecole Normale Sup\'{e}riure, 45, rue d'Ulm;
75005 Paris, France}
\begin{abstract}
The phase separation of a simple binary mixture of incompatible linear
polymers in solution is investigated using an extension of the
sedimentation equilibrium method, whereby the osmotic pressure of the
mixture is extracted from the density profiles of the inhomogeneous
mixture in a gravitational field.  In Monte-Carlo simulations the
field can be tuned to induce significant inhomogeneity, while keeping
the density profiles sufficiently smooth for the macroscopic condition
of hydrostatic equilibrium to remain applicable.  The method is
applied here for a simplified model of ideal, but mutually avoiding
polymers, which readily phase separate at relatively low
densities. The Monte-Carlo data are interpreted with the help of an
approximate bulk phase diagram calculated from a simple, second order
virial coefficient theory.  By deriving effective potentials between
polymer centres of mass, the binary mixture of polymers is
coarse-grained to a ``soft colloid'' picture reminiscent of the
Widom-Rowlinson model for incompatible atomic mixtures. This approach
significantly speeds up the simulations, and accurately reproduces the
behaviour of the full monomer resolved model.
\end{abstract}

\maketitle

\section{Introduction}

Polymer solutions segregate into polymer-rich and polymer-poor phases
when the temperature is lowered below the theta-point.  Polymer blends
nearly always demix in the melt, because the connectivity constraints
strongly reduce the entropy of mixing\cite{rubin}.  Less is known,
however, about the phase behaviour of dilute or semi-dilute polymer
solutions involving two or more polymeric species.  In particular, the
quality of the solvent is generally not the same for different
polymeric species, which may lead to competing mechanisms in the
subtle free-energy balance between entropic and energetic
contributions.  In the case of binary mixtures of incompatible
polymers in a common good solvent, excluded volume effects lead to
strong deviations from mean-field
behaviour\cite{broseta,sariban}.  In this paper we consider a
highly simplified model of mutually repelling polymer in a common
$\theta$ solvent. The system under consideration extends the familiar
Widom-Rowlinson model\cite{widom} for incompatible atomic mixtures to
polymer blends in solution. Not surprisingly, the mutual
incompatibility of the two species in the athermal model leads to
demixing above a critical density. The additional feature introduced
in the present work is the presence of an externally applied field,
specifically a gravitational field, acting on the polymers, which
induces a spatial inhomogeneity.

 The interest of the applied field is two-fold.  The first is
essentially methodological: as we have shown recently for cases of
single-component polymer solutions\cite{add2} and of di-block
copolymer solutions \cite{add3}, the density profiles characterising
the inhomogeneous distribution of polymers may be exploited to extract
the osmotic equation-of-state over a wide range of concentrations in a
single simulation\cite{biben}.  A second motivation is that applying
an external field allows us to rapidly search through phase space to
locate phase-transitions, although it must be kept in mind that adding
a field affects the phase behaviour, as demonstrated in a recent
theoretical investigations of colloid-polymer mixtures\cite{schmidt}.

 Finally, we demonstrate in this paper the merits of coarse-grained
representations of complex polymeric systems, whereby the total number
of degrees of freedom is drastically reduced by tracing out individual
monomer coordinates to determine effective interactions acting between
polymer centres of mass.  This strategy has proved very successful for
solutions of linear polymers\cite{daut,bolh,bolh2}, mixtures of asymmetric star
polymers\cite{Maye04}, di-block copolymers\cite{add3} and is extended
here to linear polymer mixtures, in an effort to map out their phase
diagram.  Our coarse-graining strategy is similar in spirit to that
successfully applied by the Klein group to surfactants and other
self-assembling systems\cite{Niel04}.

\section{Model and methodology}
Consider a binary polymer mixture of linear homopolymers with $M_A$
and $M_B$ monomers respectively.  For computational purposes the two
species are assumed to ''live'' on a cubic lattice of $L_x \times L_y
\times L_z$ sites. Monomers occupy lattice sites, and the lattice
spacing coincides with the segment length $b$.  Let $N_A$ and $N_B$ be
the total numbers of polymers of the two species.  The overall polymer
densities (coils per unit volume) are $\rho_\alpha=N_\alpha/V$, where
$V=L_xL_yL_zb^3$ (we set $b$  to $1$ in this paper). An external force field
along the $z$-direction,
acting on monomers of each species $\alpha$, and deriving from the
potential $\psi_\alpha(z)(\alpha=A$ or $B$) will induce an
inhomogeneity in the mixture characterised by two density profiles
$\rho_\alpha(z)$.  Alternatively, one may define the overall polymer
density $\rho(z)=\rho_A(z) +\rho_B(z)$ and the concentration
ratio profile:
\begin{eqnarray}
x(z)=\rho_A(z)/\rho(z)
\end{eqnarray}
The density profiles are normalised such that:
\begin{eqnarray}
\int_0^{L_z} \rho_\alpha(z)dz=n_\alpha=\frac{N_\alpha}{L_xL_y}
\end{eqnarray}
If the inhomogeneity induced by the external field varies sufficiently slowly
in space, the binary mixture obeys the hydrostatic equilibrium condition:
\begin{eqnarray}
\label{eq3}
\frac{dP(z)}{dz}=-\rho_A(z)\frac{d\psi_A(z)}{dz}
-\rho_B(z)\frac{d\psi_B(z)}{dz}
\end{eqnarray}
where $P(z)$ is the local osmotic pressure of the binary polymer
mixture in solution.  The macroscopic condition\cite{sariban} may be
shown to follow from the ``local density approximation'' (LDA) within
density functional theory of inhomogeneous fluids\cite{biben}.  If the
profiles $\rho_A(z)$ and $\rho_B(z)$ are computed in Monte-Carlo (MC)
simulations of binary mixtures subjected to the external field, $P(z)$
may be determined by integration of equation \ref{eq3}.  The
equation-of-state of the inhomogeneous mixture along an isotherm,
$P(\rho_A, \rho_B)$ or $P(\rho, x)$ may then be extracted by
eliminating the altitude $z$ between $P(z), \rho_A(z)$ and $\rho_B(z)$
(or $\rho(z)$ and $x(z)$).

 In practice, the external field is chosen
to be the gravitational field acting on the centre-of-mass of the
polymers:
\begin{eqnarray}
\label{eq4}
\psi_\alpha(z)=M_\alpha m_\alpha gz
\end{eqnarray}
where $m_\alpha$ is the buoyant mass of monomers of species $\alpha$
and g is the acceleration due to gravity.  Since gravity acts as a
confining field, it is convenient to take $L_z \rightarrow \infty$.
Although, under normal experimental conditions, gravity is
insufficient to induce significant inhomogeneity in a polymer solution,
(ultra-centrifugation would be required to induce a measurable
effect) here gravity may be tuned at will in simulations to achieve
sedimentation lengths $\zeta=k_BT/m_\alpha g$ comparable to radii of
gyration $R_{g\alpha}$, and hence a significant variation of the density
profiles.

Substitution of equation \ref{eq4} into equation \ref{eq3} and integration
leads to :
\begin{eqnarray}
\label{eq5}
P(z)/k_BT=1/\zeta_\alpha \int_z^\infty \rho_\alpha(z')dz' +1/\zeta_\beta
\int_z^\infty \rho_\beta(z')dz'.
\end{eqnarray}
The equation-of-state $P(\rho_\alpha, \rho_\beta)$ of the
heterogeneous mixture may thus be determined from the density profiles
$\rho_\alpha (z)$ and $\rho_\beta (z)$ as explained above.  In
practice MC runs are carried out for several overall concentrations,
in order to map out the complete equation-of-state. It is important to
realise that, in general, not only the overall density $\rho(z)$, but
also the concentration $x(z)$ varies with altitude, as may be easily
verified for the simple case of a binary mixture of ideal
(non-interacting) polymers.

We have applied the hydrostatic
equilibrium method to a very simple non-trivial model of a binary
polymer mixture which demixes in solution.  In the model A and B
polymers freely penetrate coils of the same species, {\it i.e.} follow
random walk statistics, while A-B pairs experience excluded volume
between their monomers {\it i.e.} behave like mutually avoiding walks.
The model is one of a family of six XYZ models, where X, Y, and Z
label the statistics of A-A, A-B and B-B interaction pairs.  Thus each
of X, Y, and Z can take one of two ``values'', I for ideal
(non-interacting) and S for self (or mutually) avoiding.  In
particular SSS refers to a binary mixture of polymers in which all
interactions are self avoiding, {\it i.e.}\ a homogeneous solution of
SAW polymers, while the model under consideration here is the ISI
model.  Note that the ISI model generalises the familiar
Widom-Rowlinson model\cite{widom} to polymer mixtures; a related block
co-polymer model, where A and B are tethered, has recently been
investigated by our group\cite{add3}.  While the latter leads to
micro-phase separation, the present binary ISI model is expected to
undergo bulk phase separation due to the incompatibility of the A and
B components. For the sake of simplicity, the following investigations
are restricted to the case of equal length polymers ($M_A=M_B=M$).
Moreover, without loss of generality, the monomer masses may be
assumed to be equal ($m_A=m_B=m$), so that $\zeta_A=\zeta_B=\zeta$.
Note that this model is athermal.  Polymers were subjected to pivot
and translational MC moves, typically $1\times 10^6 - 1\times 10^8$ MC
moves per polymer.

\section{Coarse-grained description}
Following a known strategy\cite{daut,bolh,bolh2}, one can calculate
effective interactions $v_{\alpha \beta}(r)$ between CMs of A-A, A-B
and B-B pairs, by averaging over the conformations of two polymers for
fixed values of the distance $r$ between their CMs.  Explicitly, for
an isolated pair of A and B polymers:
\begin{eqnarray}
\label{eq6}
v_{\alpha \beta}(r)=-k_BT ln P_{AB}(r)
\end{eqnarray}
where $P_{AB}(r)$ is the probability (averaged over all random walk
configurations of the two polymers) that there is no overlap between
monomers of A and monomers of B when their CMs are held at a distance
$r$.  Obviously $v_{AA}(r)=v_{BB}(r)=0$ for the ISI model, since
like-species overlaps are allowed.  Effective pair potentials like
$v_{AB}$ in eq.\ \ref{eq6} are easily computed in MC simulation, as
explained on more detail in the references\cite{bolh,bolh2}.
\begin{figure}[!htp]
\begin{center}
\includegraphics[width=3.25in,angle=0]{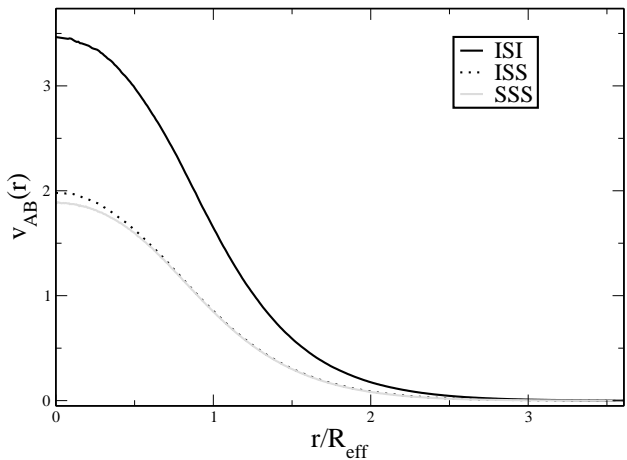}
\caption{Effective potential $v_{AB}(r)$, in units of $k_BT$, between
 the centres of mass of two polymers for ISI, ISS and SSS models.
 These potentials were generated at zero density using Monte Carlo
 simulations of two $L=500$ polymers on a simple cubic lattice.}
\label{fig1}
\end{center}
\end{figure}
Examples are shown in figure \ref{fig1} for the ISI, ISS and SSS
models; since the two species are of equal length, the SSS model is
equivalent to a one-component system of self-avoiding walk (SAW)
polymers, which has been thoroughly
investigated\cite{bolh,bolh2,pelis}.  While the potentials $v_{AB}(r)$
for the SSS and ISS models are very close, the effective potential for
the ISI model is considerably more repulsive,with a value at full
overlap, $v_{AB}(r=0)\simeq 3.45k_BT$.  This enhancement of the
effective repulsion may be easily understood by noting that the A and
B coils are more compact (their radii of gyration are those of ideal
chains, which scale as $R_g\sim M^{0.5}$, as compared to chains with
SAW internal statistics, which scale as $R_g\sim M^{0.59}$). Hence the
overlap probability is higher ({\it i.e.}\ $P_{AB}(r)$ is lower) in
the former compared to the latter case, for any given $r/R_{eff}$
where the effective radius is defined by
\begin{eqnarray}
\label{eq7}
R{^2}_{eff}=\frac{1}{2}(R_{gA}{^2} +R_{gB}{^2}).
\end{eqnarray}
The second virial coefficient for the ISI model may be calculated from
$v_{AB}(r)$ according to:
\begin{eqnarray}
\label{eq8}
B_{AB}=-2\pi \int_0^\infty(e^{-v_{AB}(r)/k_BT}-1)r^2dr
\end{eqnarray}
and turns out to be $B_{AB}\simeq 9.5 R_g^3$ for our $L=500$ ISI
model lattice polymers.

It has been shown\cite{bolh} that the effective pair potential between
linear SAW polymers varies significantly as the polymer density is
increased from the infinite dilution limit, particularly so beyond the
overlap density $\rho^*=3/4\pi R_g^3$.  This effect is expected to be
less important for the ISI model, since it will be shown that phase
separation, which leads to a dramatic reduction of the number of A-B
contacts, sets in for $\rho$ greater than $\sim \frac12 \rho^*$.

\section{Equation-of-state of the symmetric mixture}
\begin{figure}[!htp]
\begin{center}
\includegraphics[width=3.25in,angle=0]{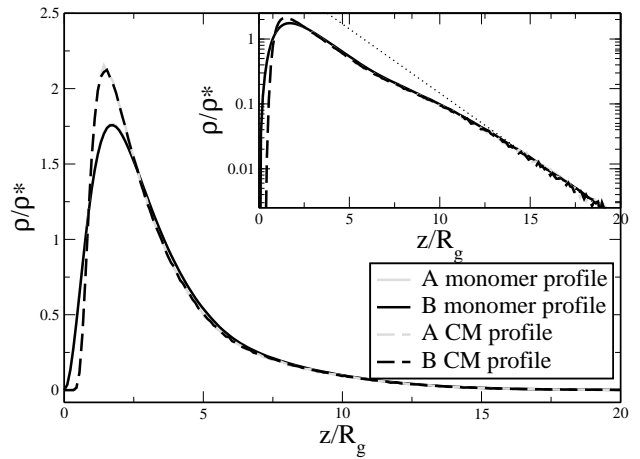}
\caption{The centre of mass and monomer density profiles for the
symmetric equimolar ISI model, with the same profiles on a logarithmic
scale shown as an insert. These profiles were generated from
simulations of $1600$ polymers of length $L=500$ in a box with a base
of $L_x=L_y=200$ that is open in the positive $z$ direction, along
which the density profiles are plotted.  The dashed line on the
logarithmic scale (inset) shows the exponential behaviour of ideal polymers
$\rho(z) \sim \exp(-z/\zeta)$, with the gravitational length $\zeta =
2.19R_g$.}
\label{fig2}
\end{center}
\end{figure}
In order to determine the full equation-of-state $P(\rho_A,\rho_B)$ or
$P(\rho,x)$ by the hydrostatic equilibrium method presented in section
2, simulation of the ISI model must be carried out on systems
involving several overall compositions.  In the equimolar case, the
ISI model is fully symmetric with respect to the A and B species, so
that the local concentration $x(z)$ remains constant (equal to
$x=\frac12$) at all altitudes.  Examples of MC-generated monomer and
CM density profiles $\rho_A(z)$ and $\rho_B(z)$ of the equimolar ISI
model are shown in figure \ref{fig2}. As expected, the density profiles of
the two species are identical within statistical errors.  The two
profiles exhibit a depletion zone (which is more pronounced for the CM
profile) near the bottom, followed by a peak at $z \sim 2R_g$, which
is again more pronounced for the CM profile. At higher altitudes the
profiles go over to the exponential barometric behaviour of ideal
particles, as clearly seen in the inset. The intermediate behaviour
signals phase separation (as discussed below). Beyond the peak,
the CM and monomer profiles are virtually indistinguishable. Only the
latter part is used in the inversion procedure to extract
the osmotic equation-of-state, since the rapid variation at low
altitudes is incompatible with the LDA, and hence with the macroscopic
condition of hydrostatic equilibrium embodied in eq.\ \ref{eq5}.
\begin{figure}[!htp]
\begin{center}
\includegraphics[width=3.25in,angle=0]{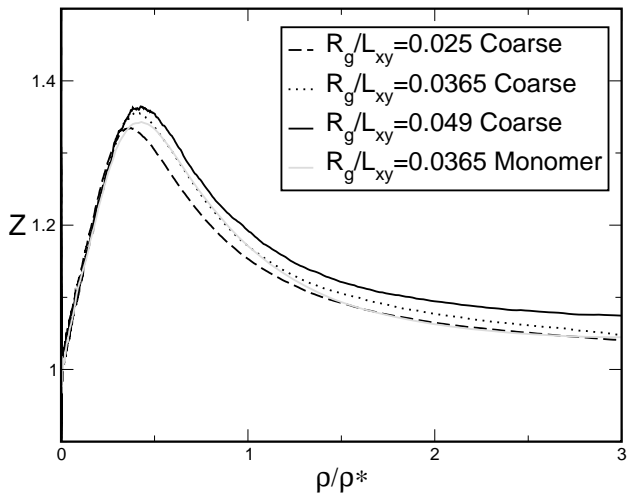}
\caption{The equation-of-state $Z=\beta P/\rho$ for different box
sizes, denoted by $R_g/L_{XY}$, calculated by coarse grained
simulations and compared, at $R_g/L_{XY}=0.0365$, to an equivalent
fully resolved monomer-level simulation }
\label{fig3}
\end{center}
\end{figure}
\begin{figure}[!htp]
\begin{center}
\includegraphics[width=3.25in,angle=0]{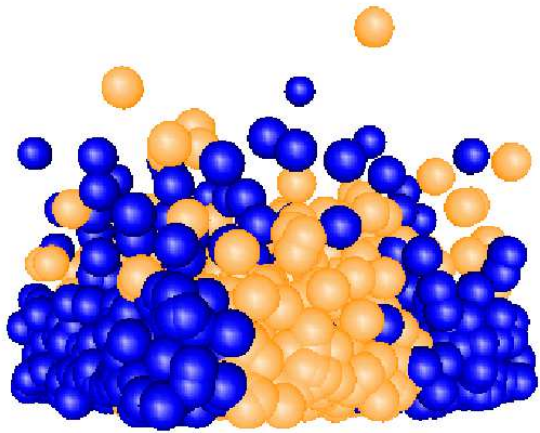}
\caption{Snapshot of a configuration for $N_A=N_B=800$ polymers,
showing the CM of each as a sphere of radius $R_g$.  The sedimentation
lengths are $\zeta_A=\zeta_B=2.11R_g$, and the open box has a base of
length $L_X=L_Y=200$. Note that the existence of two interfaces is due
to the use of periodic boundary conditions in the x and y directions.}
\label{fig4}
\end{center}
\end{figure}
The resulting equation-of-state $Z=P/\rho k_BT$ is plotted in
figure \ref{fig3}.  After an initial linear increase, $Z$ goes through a
maximum around $\rho/\rho^* \lesssim 0.5$, before decreasing and
apparently saturating at a value above one at high densities.  This
unusual behaviour may be traced back to to the phase separation, which
is evident in the snapshot of a typical configuration shown in fig
\ref{fig4}, and which will be analysed in more detail in section
5.   The existence of phase
boundaries leads one to expect a substantial dependence of the
measured equation-of-state on the size of the simulation box.

To
illustrate this size dependence, we have carried out MC simulations of
the coarse-grained representation of the ISI model, based on the
effective potential $v_{AB}(r)$ introduced in section 3 (fig
\ref{fig1}) for three values of the ratio $R_g/(L_{XY})$ (where
$L_{XY}=L_X=L_Y$ is the length of the square base of the simulation
cell).  These were carried out for $N_A=N_B=800-2000$ coarse-grained
particles, with a sedimentation length $\zeta=2R_g$. Such simulations
are typically two orders of magnitude faster than those based on the
full monomer-level ISI model for the same number of polymers. For
$R_g/(L_{XY})=0.0365$, the agreement between the equations of state
calculated with the two representations is seen to be satisfactory in
figure \ref{fig3}. The small discrepancies are probably due to the fact
that we have neglected the density dependence of the effective
potential $v_{AB}(r)$, for which we have used the zero density result
shown in figure \ref{fig1}.  The simulation data for the coarse-grained
model plotted in figure \ref{fig3} clearly illustrate the size dependence
at high densities, {\it i.e.} in the region of phase coexistence. This is
expected since the
smaller the system, the larger the interfacial contribution to the
equation-of-state; the former will become negligible only in the
thermodynamic limit ($R_g/(L_{XY})\rightarrow0$).

\section{Interpretation of the simulation data}
The equation-of-state data of figure \ref{fig3} can be understood in
terms of an underlying phase separation, as suggested by fig
\ref{fig4}.  The phase diagram of the ISI model can be calculated, at
least approximately, from the virial expansion of the free energy.
The free energy per unit volume can be expanded as:
\begin{eqnarray}
\label{eq9}
f=\frac{F}{Vk_BT}=f_A^{id}+f_B^{id}+2x(1-x)B_{AB}\rho^2+{\cal O}(\rho^3)
\end{eqnarray}
where $f^{id}_\alpha=\rho_\alpha(ln(\Lambda_\alpha^3 \rho_\alpha)-1)$
and $B_{AB}$ is the second virial coefficient of equation
(\ref{eq8}). Differentiation of (\ref{eq9}) with respect to $\rho$ yields:
\begin{eqnarray}
\label{eq10}
Z=\frac{P}{\rho k_BT}=df/d\rho-f/\rho=1+2x(1-x)B_{AB}\rho+{\cal O}(\rho^2).   
\end{eqnarray}
Above a critical density $\rho_c$, the free energy (\ref{eq9}) leads to
phase separation; by symmetry the critical concentration $c_c=1/2$,
while $\rho_c=1/B_{AB}$.  Inserting the numerical value
$B_{AB}=9.5R_g^3$ reported in section 3, one finds
$\rho_c/\rho^*=0.44$.  Following an analysis similar to one presented
in ref.\ \cite{louis} one can calculate the spinodal curve analytically:
\begin{eqnarray}
\label{eq11}
\rho_s(x)=\frac{1}{2B_{AB}\sqrt{x(1-x)}}.
\end{eqnarray}
\begin{figure}[!htp]
\begin{center}
\includegraphics[width=3.25in,angle=0]{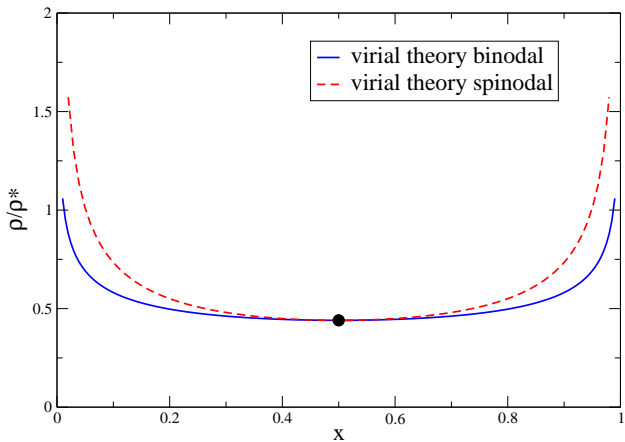}
\caption{Spinodal and binodal demixing lines for the ISI model,
calculated from second virial coefficient expansion of the free energy
described in Eqs.~\protect\ref{eq9}~-~\protect\ref{eq11}.}
\label{fig5}
\end{center}
\end{figure}
\begin{figure}[!htp]
\begin{center}
\includegraphics[width=3.25in,angle=0]{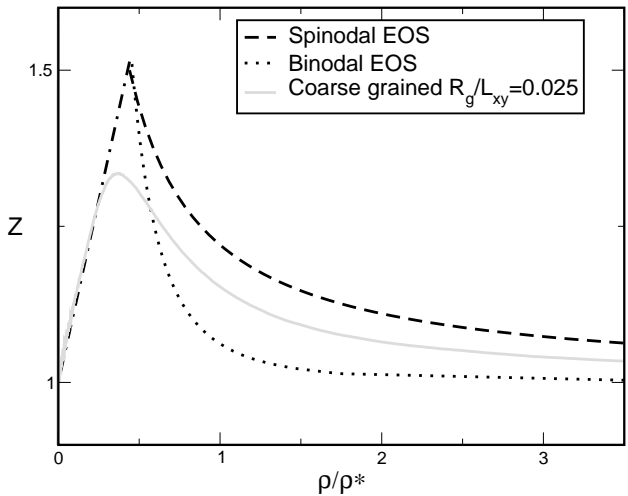}
\caption{The equation-of-state from the largest coarse-grained system
simulations is compared to theoretical curves, calculated from the
binodal and the spinodal lines of a second virial expansion of the
free energy described in Eqs.~\protect\ref{eq9}~-~\protect\ref{eq11}.}
\label{fig6}
\end{center}
\end{figure}
The binodal is easily calculated numerically, and results are
summarised in figure \ref{fig5}; segregation of the two species is seen to
be nearly complete for $\rho/\rho^* \sim 1$.  The equation-of-state
data of figure \ref{fig3} may now be ``read'' as follows.  As the
altitude $z$ decreases in the polymer sediment, the local density
$\rho(z)$ increases ({\it c.f.} figure \ref{fig2}).  Up to
$\rho/\rho^*\lesssim 0.4$ the system remains in the one phase region of
the phase diagram in figure \ref{fig5}, and the equation-of-state Z
increases linearity with $\rho/\rho^*$ according to (\ref{eq10}); the
slope of the linear portion of the MC-generated equation-of-state
agrees well with that predicted by (\ref{eq10}) (with $B_{AB}\simeq
9.5R_g^3$). When the local density $\rho(z)$ increases beyond the
critical density $\rho_c/\rho^* \simeq 0.44$, the system enters the
phase-coexistence region bounded by the binodal curve and separates
symmetrically into two mixtures with compositions $x$ and $1-x$ which
may be read from figure \ref{fig5}.  The excess contribution to the
pressure of the coexisting phases decreases with density, because the
number of A-B contacts drops sharply in the A and B-rich phases as the
overall density rises.  At densities $\rho >\rho^*$, the nearly
complete segregation means that coexisting phases behave practically
as ideal cases of A or B polymers.  Hence, for sufficiently large
systems such that the interfacial contribution to the local stress
is negligible compared to the bulk contributions,
the equation-of-state $Z$ must return to its ideal gas value 1 for
$\rho\gg\rho^*$.

The theoretical equation-of-state is compared in
figure \ref{fig6} to the MC data for the largest system.  The agreement
is seen to be only qualitative.  The theoretical curve has a cusp
point (rather than a rounded maximum) at the density where phase
separation starts.  In the simulations, the system may enter the
metastable region comprised between the binodal and spinodal curves in
figure \ref{fig5}.  In that case the mixture may reach higher densities
before segregating.  The equation-of-state calculated on the assumption
that the mixture separates into two phases along the spinodal curve is
also shown in figure \ref{fig6}.  It is gratifying to realise that the
simulation data lie between the binodal and spinodal
equations-of-state, which constitute two extreme limits of the actual
scenario.

\begin{figure}[!htp]
\begin{center}
\includegraphics[width=3.25in,angle=0]{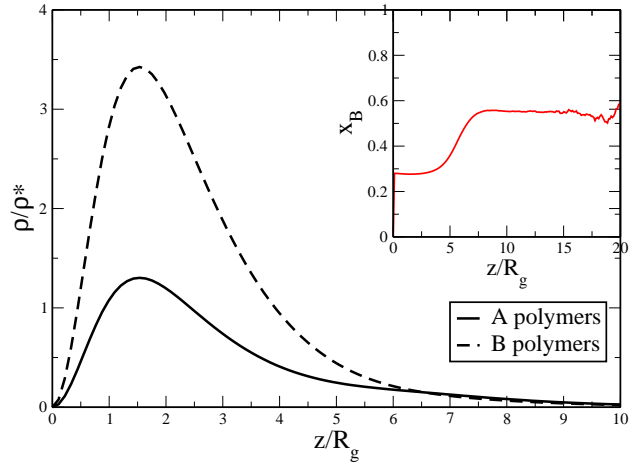}
\caption{Density and concentration profiles $\rho_A(z)$, $\rho_B(z)$,
$x(z)$ (inset) of a 70:30 mixture of ISI polymers.}
\label{fig7}
\end{center}
\end{figure}
In the more general case where the overall composition of the binary
polymer solution $x\neq 0.5$ ({\it i.e.\ }$N_A \neq N_B$), the local
concentration ratio, $x(z)$ will no longer be constant, and the extraction
of $Z(\rho,x)$ from equation (\ref{eq5}) requires several simulations
of systems with different compositions.  No attempt is made at a full
investigation in the present paper, but the situation is illustrated
in figure \ref{fig7}, where density and concentration profiles obtained
from MC simulations of the full monomer description of a system with
$N_A=480$ and $N_B=1120$ ($x=0.3$) polymers of $L_A=L_B=500$ in an
open square based box of side length $L_{XY}=200$ are plotted.  The
concentration profile $x(z)$ is seen to vary substantially at
altitudes where phase separation is expected to take place.

In this case, as well as for lower overall concentration fractions of
A particles, we observe the formation of a vertical, cylindrical
interface, rather than two planar interfaces as in figure \ref{fig4}.
This observation can be understood as the consequence of a
minimisation of the area over which unfavourable A-B contacts take
place.  Assuming that the distributions $\rho_A(z)$ and $\rho_B(z)$ of
the two species are identical irrespective of the geometry of the
interface, and that the height H of the two types of interfaces is the
same, and noting that the radius r of the cylinder is related to the
concentration $x$ by $x=\pi(r/L_{XY})^2$, one finds that the areas of
the two planar and cylindrical interfaces are identical when
$x=1/\pi$; below this value, the cylindrical interface will be
preferred.  We explicitly checked this by performing simulations for
$N_A+N_B = 1600$ polymers at $x=0.1,0.2,0.3,0.4$ and $0.5$. As
expected, for the lowest three concentration ratios we find
cylindrical interfaces, and for the two higher concentrations we find
planar interfaces.

\section{Conclusion}
We have extended the sedimentation equilibrium method for the
computation of the osmotic equation-of-state of polymer solutions to
the case of binary polymer mixtures undergoing phase separation.  The
method was applied to the athermal ISI model, where polymer species A
and B (assumed here to be of the same length) behave like ideal
(random walk) polymers, but are mutually avoiding, {\it i.e.} monomers
of different species cannot occupy the same lattice sites.  This
incompatibility leads to segregation for overall polymer densities
higher than about one half of the overlap density $\rho^*$.  This
segregation leads to a somewhat unusual variation of the osmotic
pressure with density, since outside the interfacial region, the
mixture behaves as an ideal solution both in the low and high density
limits.

 The MC generated equation-of-state agrees well with the data obtained
from a coarse-grained representation based on an effective interaction
between the CMs of A-B pairs, and agrees semi-quantitatively with the
predictions of a simple analytic theory based on a second virial
expansion.  The cusp in the equation-of-state predicted by the latter
is rounded into a maximum in the MC data due to finite size effects.
The present investigation suggests that the phase behaviour of
segregating mixtures can be extracted from a measurement of the
density profile of the inhomogeneous mixture in an external field.
Although our model is highly oversimplified, we have demonstrated that
our coarse-graining strategy can be applied in a straight-forward
fashion to the demixing of polymer solutions.  We plan to extend this
methodology to more realistic polymer models, and to grand canonical
ensemble simulations.

\acknowledgements
C.I Addison wishes to thank the EPSRC for a studentship, and A.A. Louis
is grateful to the Royal Society of London for a University Research
Fellowship.  J.-P. Hansen would like to express his gratitude to Michael Klein
for inspiration and unfaltering friendship over many years.

\end{document}